\documentclass[12pt]{iopart}
\usepackage{graphicx}

\newcommand{\IPT}{${\cal I}\pi\tau$}

\begin{document}

\title[Geometric phases in a ring]{Geometric phases of scattering states in a ring geometry: adiabatic pumping in mesoscopic devices}

\author{Huan-Qiang Zhou\dag\ 
\footnote[3]{To whom correspondence should be addressed (hqz@maths.uq.edu.au)}, 
Urban Lundin\dag , and Sam Young Cho\dag}
\address{\dag\ School of Physical Sciences, University of Queensland,
	     Brisbane Qld 4072, Australia}

\begin{abstract}
Geometric phases of scattering states in a ring geometry are studied based  
on a variant of the adiabatic theorem. 
Three time scales, i.e.,  
the adiabatic period, the system time and the dwell time, 
associated with adiabatic scattering in a ring geometry 
plays a crucial role in determining geometric phases, 
in contrast to only two time scales, i.e., the adiabatic period 
and the dwell time, in an open system. 
We derive a formula connecting 
the gauge invariant geometric phases acquired by time-reversed scattering 
states and the circulating (pumping) current. 
A numerical calculation shows that the effect of the geometric phases is 
observable in a nanoscale electronic device.
\end{abstract}
\submitto{\JPCM}
\maketitle

\section{Introduction} 
The study of geometric phases continues to be an intriguing subject. 
Since Berry's original work~\cite{berry}, 
various generalizations have been proposed~\cite{sw89,jones00} 
to adapt to different physical applications in diverse fields. 
In Ref.~\cite{zhou1} the concept of geometric phases for scattering 
states was introduced which allows to give a geometric interpretation of 
the formalism developed by Brouwer~\cite{b98} 
for quantum adiabatic pumping in open systems. 
Specifically, Brouwer presented a compact
formula for the pumped charge (current) in terms of 
the parametric derivatives of the time-dependent scattering matrix subjected 
to the oscillating potential, which has been identified
as the geometric phase accompanying the scattering state 
associated with the time-reversed Hamiltonian. 
Quantum adiabatic pumping in open systems is subject to 
intense study~\cite{b98,mb02,zhou1,zhou2} 
due to the observation of 
adiabatic electron transport through a quantum dot subject to slow cyclic 
variation of gate voltages~\cite{switkes}. 

An intriguing question has been raised recently about 
the possibility to generate a
circulating (pumping) current in a ring geometry by 
adiabatically varying external 
parameters~\cite{cohen03,mb03}. In Ref.~\cite{cohen03}, Cohen considered 
open systems as some subtle limit of closed systems and demonstrated
that it is possible to reproduce the formulas 
by Landauer~\cite{landauer} and B\"uttiker,
Pr\`etre and Thomas~\cite{btp} from the Kubo formula 
whereas Moskalets and B\"uttiker~\cite{mb03} 
examined this problem in the
slow frequency limit of the Floquet scattering theory. 
These differences in approaches mean that the results are different, although 
they aimed at describing the same system. 
Such an unsatisfying situation calls for a 
thorough analysis of the problem.
In this Letter, we address this issue from gauge field and
geometric phase perspectives. 
We derive a compact formula connecting the 
circulating (pumping) current~\cite{mailly93} and 
the gauge invariant geometric phases acquired by 
time-reversed scattering states. This is achieved by a purely geometric 
argument, based on the fact that pumped charge is 
additive for two consecutive pumping cycles. 
A generic feature of pumping in a ring 
geometry is that the momentum is time-dependent and so are the 
discrete energy levels.  
The closed system is different from the open in 
another aspect. In an open system there are only two different time scales, 
whereas in the closed ring geometry three different time scales appear, 
the adiabatic period, the system time and the dwell time. 
The physics behind this is the following. In the 
adiabatic limit, electrons are in an 
instantaneous scattering state. A small
deformation in the scattering potential 
results in a small increment of geometric phases,
which electrons can pick up at the dwell time scale. 
The dwell time being the time it takes for an electron to complete 
one scattering event. Since the electrons travel in a ring geometry, they 
have to obey the periodic boundary condition to which they adjust at 
the system time scale, $\tau_s$. 
All electronic wave functions are extended over the entire ring
thus inducing a circulating current. The latter
is experimentally measurable in terms of a SQUID that measures the 
magnetic field produced by the circulating current. 
 
\section{The system}
\begin{figure}[hbt]
\begin{center}
\includegraphics[scale=1.0]{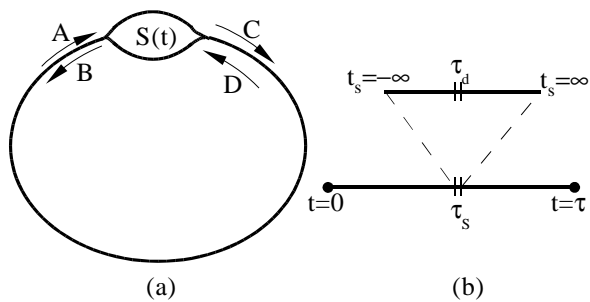}
\end{center}
\caption{\label{setup:fig}
(a) shows the setup with a time dependent scatterer, 
defined by its scattering matrix, $S(t)$. Incident, $A$, and reflected wave, 
$B$. 
(b) shows the time scales relevant to the problem. Scattering takes place 
on $\tau_s$, the dwell time (the time it takes for the scattering event) is 
$\tau_d$. $t_s$ designates the time passing for the scattering event, it is 
only valid on the $\tau_s$-scale. 
The time it takes for the wave function to adjust the boundary 
conditions to the scatterer, of the order of $\tau_s$, must be shorter than 
the adiabatic changes in the scatterer, happening on the scale $\tau$.}
\end{figure}

Consider a quantum dot with two leads attached such that 
the two single-channel leads are bent back to 
form a ring [see Fig.~\ref{setup:fig}(a)]. 
Suppose the dot is characterized by a scattering 
matrix $S(t)$ which 
depends on time $t$ via a set of independent external 
parameters $X \equiv (X^1, \cdots, X^{\nu}, \cdots, X^p)$ 
oscillating slowly 
with frequency $\omega$. For a spin-independent scatterer, $S(t)$ is a 
$2 \times 2$ matrix. Define the row vectors 
${\mathbf n}^L \equiv (r, t)$, and 
${\mathbf n}^R \equiv (t',r')$,
the unitarity of the scattering matrix implies that they are orthonormal 
${\mathbf n}^{\alpha} \cdot 
{\mathbf n}^{\beta}=\delta _{\alpha\beta}$, where 
$\alpha,\beta=L,R.$
From the boundary condition of the matching of the wavefunction
for a ring geometry,
the dispersion equation is written in terms of scattering matrices,
$\det (S S_w -1)=0$ with $S_w = \exp (ikL) \sigma ^x$, 
where $\sigma ^x$ is the Pauli matrix. 
$S_w$ is the scattering matrix describing the remaining part
of the ring except the dot. 
The energy spectrum is then discrete in the ring geometry. 
The adiabaticity requires that the frequency
$\omega$ is small enough compared with the (averaged) level spacing 
$\Delta$. On the other hand, for the instantaneous scattering matrix to 
make sense, it is necessary that the system time 
$\tau_s \equiv L/v$ ($v$- electron velocity) is much greater than the 
dwell time $\tau_d$\cite{dtime}
during which electrons scatter off the scatterer. 
That is, in the ring geometry, there are three different time scales:
the period $\tau =2 \pi /\omega$, the system time $\tau _s = 
\hbar \Delta ^{-1}$, 
and the dwell time, $\tau _d$. In the scattering approach to adiabatic 
quantum pumping in the ring geometry, 
we assume $\tau >> \tau _s >> \tau _d$.
This condition ensures that it is legitimate to speak of instantaneous
scattering states, consistent with Heisenberg 
uncertainty principle~\cite{thirring}.  
We illustrate this in Fig.~\ref{setup:fig}(b) and notice that it is
different from the adiabaticity condition discussed in Ref.~\cite{mb03}. 
Generically, the allowed (discrete) $k$'s, which in turn determine the
discrete energy levels, are time-dependent due to the fact that the 
external parameters defining the scatterer 
$X \equiv (X^1, \cdots, X^{\nu}, \cdots, X^p)$ are changing with time and 
the scattering matrix are decisive for the dispersion, 
a feature different from scattering in open systems.

\begin{figure}[hbt]
\begin{center}
\includegraphics[angle=270,width=\columnwidth]{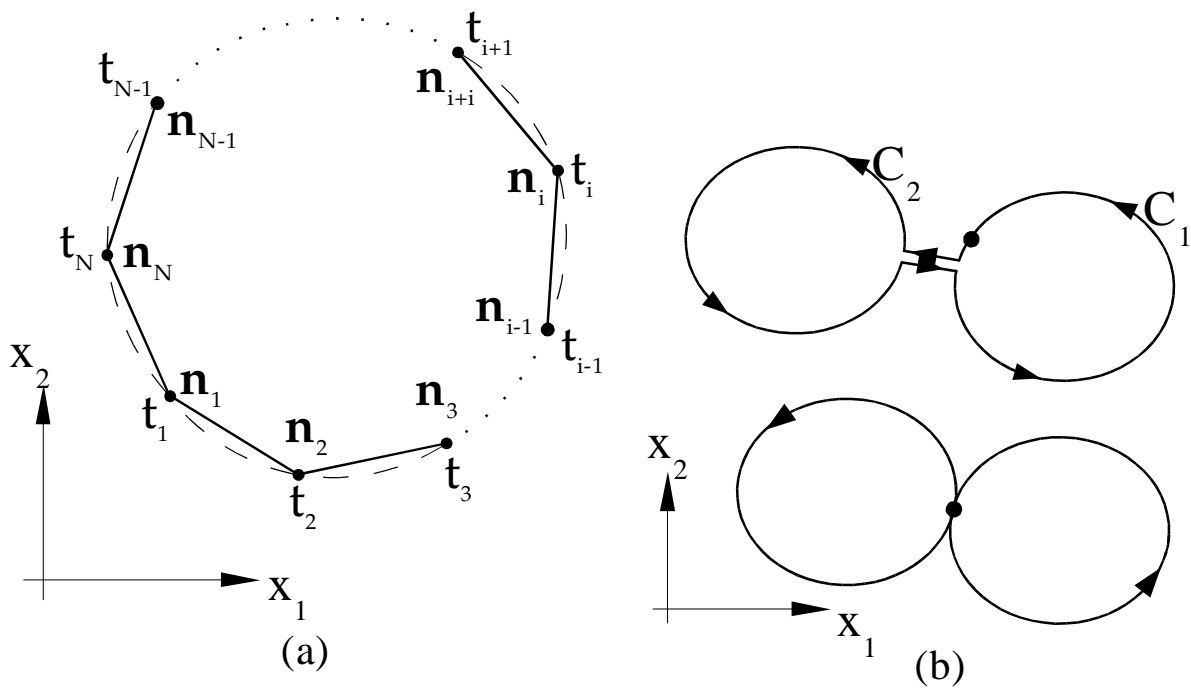}
\end{center}
\caption{\label{contour:fig}
(a) displays the discretization of the adiabatic loop. 
Doing this, we can ensure that the scattering matrix formalism gives a correct 
description and distinguish the time scales associated with the theory. 
Here, the entire interval $[0,\tau]$ is divided into $N$ pieces. Each 
subinterval $[t_i,t_{i+1}]$ is of the same order of magnitude as the system 
time $\tau_s$. At each instant, $t_i$, two new vectors $n_i^L$ and $n_i^R$ 
are defined in terms of the instantaneous $S$-matrix. 
(b) displays the property that the geometric phase is additive, and the 
pumped charge being a function of the geometric phase has to be additive. 
The lower contour 
displays a special case in which two cycles share the same initial 
configuration, represented by a dot, whereas the upper contour 
shows two separate cycles joined by two adjacent but opposite lines.}
\end{figure}

{\it Discretized approach and the continuous limit.} 
Suppose the system undergoes an adiabatic cycle with the period $\tau$, 
which is characterized by a closed loop in parameter space. 
Suppose that initially an electron is in a given discrete energy
level, the adiabaticity ensures that it remains in the same
level during the entire period.  To take into 
account the time scales properly, we adopt a discretized approach, 
i.e., the entire interval $[0,\tau]$ is divided into $N$ pieces. 
Correspondingly, the 
closed loop may be regarded approximately as a polygon 
[see Fig.~\ref{contour:fig}(a)]. 
Here we stress that each subinterval $[t_i, t_{i+1}]$ is at 
the same order of magnitude as 
the system time $\tau _s$. At each instant 
$t_i$, we have two row vectors ${\bf n}^L_i$ and ${\bf n}^R_i$. 
Since the argument is parallel, we drop the superscripts $L$ and $R$ for
brevity. 
For two consecutive instants $t_i$ and $t_{i+1}$, we may define 
the relative phase between the two vectors ${\bf n}_i$ and
${\mathbf n}_{i+1}$ by 
$ \exp ({i \phi_i}) = 
{\mathbf n_i}^* \cdot {\mathbf n}_{i+1}/ |{\mathbf n_i} 
\cdot {\mathbf n}^*_{i+1}|$. That is, 
$\phi_i = {\mathrm Im} \ln {\mathbf n_i}^* \cdot {\mathbf n}_{i+1}$.
Therefore we have the total phase for the closed loop (polygon) 
\begin{equation}
\phi \equiv \sum^N _{i =1}\phi _i =  \mathrm{Im} \ln \prod ^N_{i =1}
{\bf n_i}^* \cdot {\bf n}_{i+1}.
\label{phi:eq}
\end{equation} 
That is, both ${\bf n}^L$ and ${\bf n}^R$ acquire a geometric phase 
$\phi ^L$ and $\phi ^R$, respective, if 
the system returns to the initial 
configuration after one adiabatic cycle. Indeed, Eq.~(\ref{phi:eq}) 
is invariant under the (local) gauge transformation 
${\bf n}'_i = e^{i \alpha_i} {\bf n}_i$, which 
arises from the fact that
the absolute phase is not observable in quantum mechanics. 
In the continuum limit, we have
\begin{equation}
\phi =  \mathrm{Im} \oint {\bf n}^*\cdot d{\bf n}.
\end{equation}
This amounts to the statement that the gauge potential
$A$ is $\mathrm{Im}({\bf n}^*\cdot d{\bf n})$ up to an (integrable) 
immaterial 
term. 
Indeed, in the adiabatic limit, 
it is reasonable to assume that the gauge potential $A$ 
{\it only} depends on 
${\bf n}, d{\bf n}, k, dk$ and $L$. From this we see
that the most general 
gauge potential we can construct takes the form 
$A={\bf n}^*\cdot d{\bf n}+g(kL) dk L$ with $g$ an arbitrary function,
if we require that $A$ transforms as 
$$ A' = id \alpha + A,$$
under the gauge transformation. Here the second term proportional to $dk
L$ comes from the dependence of discrete $k$'s on time. 
Below we will see that
$g=i/4$. Although the gauge potential $A$ is formally identical
to its counterpart in the case of open systems when expressed
in terms of the row vector ${\bf n}$, the explicit dependence of
${\bf n}$ on $k$ which varies with time characterizes the difference.

Similarly, we may define two column vectors 
${\hat {\mathbf n}}^L \equiv (r, t')$, and 
${\hat {\mathbf n}}^R \equiv (t, r')$, which are
orthonormal due to the unitarity of the $S$ matrix.  Repeating the above
argument, we conclude that  
${\hat {\bf n}}^L$ and ${\hat {\bf n}}^R$ acquire, respectively,
geometric phases 
${\hat {\phi}} ^L$ and ${\hat {\phi}} ^R$, which take the form,
${\hat {\phi}} =  {\rm Im} \oint {\hat {\bf n}}^*\cdot d{\hat {\bf n}}$.
Here the hat denotes
the time-reversal operation since the row vectors and column vectors are
connected via time reversal operation.

{\it Consistency with the periodic boundary condition.} 
The above discussion indicates that there is a gauge group 
$U_L(1) \times U_R(1)$, which is necessary to accommodate 
the geometric phases $\phi _L$ and $\phi _R$ acquired by 
the two row vectors ${\bf n}_L$ and ${\bf n}_R$, respectively. 
As we see, this results from the gauge freedom associated with
the incident waves.  In contrast 
to adiabatic pumping in open systems, one has to keep in mind that the 
scattering wave functions should satisfy the periodic boundary 
condition 
in the ring geometry. If we 
set $\Psi_L(x) = A \exp (ikx) + B \exp (-ikx)$ 
for $ -L/2 \leq x \leq 0$ and
$\Psi_R(x) = C (ikx) + D\exp (-ikx)$ for $ 0 \leq x \leq L/2$, then the 
incident waves and the scattered waves are connected via 
$B = A r + D t'$ and $ C= At + D r'$. 
On the other hand, the period boundary condition requires
that $B=D \exp (-ikL)$, and $C=A\exp (-ikL)$. 
Besides the dispersion equation which results in 
discrete energy spectrum, 
we have
$D=A r/ (\exp (-ikL)-t')$. Formally, we 
write the scattering state $\psi$ 
as $\psi = A \psi _L + D \psi _R$ with 
$\psi _L = \exp (ikx) + r\exp (-ikx) +t \exp (ikx)$ and 
$\psi _R = \exp (-ikx) + r'\exp (ikx) +t' \exp (-ikx)$. 
Then the $U_L(1) \times U_R(1)$ gauge transformation takes the form 
$\psi _{L (R)} \rightarrow \exp (i \alpha _{L(R)}) \psi _{L (R)}$, 
which induces 
transformations on $A$ and $D$: 
$A \rightarrow \exp  (-i \alpha _L) A$ and 
$D \rightarrow \exp (-i \alpha _R) D$. Then we see that under such a
gauge transformation the dispersion equation 
remains the same, but an extra phase factor 
$\exp  (-i(\alpha _L - \alpha _R))$ 
appears in the relation between 
$A$ and $D$. With $\psi_L$ and $\psi_R$ as the local basis, the time
evolution governed by the Schr\"odinger equation induces parallel
transport~\cite{pt}. Taking into account the causality condition which
states that scattered waves appear {\it only after} 
incident waves hit the 
scatterer, we may derive the expression of the gauge potential $A$
(where $g$ is determined to be i/4), 
resulting in the geometric phases $\phi _L$ and $\phi _R$.  
When the system is brought back to the initial 
configuration, the scattering 
states $\psi _L$ and $\psi _R$ acquire, respectively, geometric phases 
$\phi _L$ and $\phi _R$.  
Now we may picturize the observable effect of geometric 
phases as follows. Suppose the system undergoes an adiabatic cycle. 
During this process, at each instant, electrons are in an instantaneous 
scattering state. When electrons feel a slow deformation in the 
scattering potential, they pick up a small increment of geometric 
phases at the dwell time $\tau_d$ scale, and then take time at the 
system time $\tau_s$ scale to redistribute themselves to maintain the 
periodic boundary conditions, thus inducing the circulating current in 
the pumping experiment setup. 

In the above we focused on scattering geometric phases acquired  by
scattering states. However, in the pumping setup, physical observables
are connected with pumping geometric phases associated with
time-reversed scattering states, a fact 
already known for pumping in open
systems. In fact, there is another gauge group
$U_L(1) \times U_R(1)$ associated with the gauge freedom of the
scattered waves, which 
accommodates 
the geometric phases ${\hat {\phi}} _L$ and ${\hat {\phi}} _R$.

{\it Connection between physical observables and  geometric phases.} 
Now let us establish the connection between the geometric phases 
${\hat \phi}_L$ and ${\hat \phi}_R$ and the
physical observables which are the charge $Q_a$ accumulated inside 
the dot and the charge $Q_c$ circulating in the ring during the entire 
pumping period $\tau$ for a given occupied level. 
Suppose $Q^L$ and $Q^R$ 
are, respective, the charges pumped from 
the dot to the left and right sides.
Obviously, we have $Q_a = -(Q^L +Q^R)$. As for $Q_c$, one may expect
that it is proportional to $(Q^L-Q^R)$, i.e.,  .
$Q_c = \xi (Q^L-Q^R)$ with $\xi$ some undetermined
constant. To determine $\xi$, we consider the special case in
which no charge is accumulated inside the dot. In such a case,
we have $Q^L =-Q^R=Q$. That is, the amount of charge Q is pumped from
the left side to the right side of the dot.  Then we have
$\xi=1/2$.

As observables, $Q^L$ and $Q^R$ must be gauge invariant, 
so they must be 
some functions of the geometric phases
${\hat {\phi}}^L$ and ${\hat {\phi}}^R$. 
Since the left and right sides are 
independent, we have $Q=f({\hat \phi})$ 
for a given closed loop (here we 
drop the superscripts for brevity.). 
To determine the function $f$, we 
notice that the charge is additive. 
That is, if we consider a closed loop 
consisting of two consecutive closed loops $C_1$ and $C_2$ 
[see the lower contour in Fig.~\ref{contour:fig}(b)], then we have
$Q_C = Q_{C_1} + Q_{C_2}$. On the other hand, the geometric
phase is Abelian, implying that 
${\hat \phi}_C = {\hat \phi}_{C_1} + {\hat \phi}_{C_2}$.  That is, 
$f({\hat \phi}_{C_1} + {\hat \phi}_{C_2}) = f({\hat \phi}_{C_1})+ 
f({\hat \phi}_{C_2})$. 
>From this we see that the function must be linear, 
i.e., $Q = c {\hat \phi}$ with $c$ an undetermined constant. 
Combining this with the Friedel sum rule which states that charge 
accumulated inside the scatterer follows $\delta Q_a = e/ (2\pi i) \delta
{\rm ln } ({\rm det} S)$, we have $c= -1/ (2 \pi)$.  Actually, the
argument is applicable to any two arbitrary loops as illustrated in
the upper contour in Fig.~\ref{contour:fig}(b). 

In the context of adiabatic pumping, we are interested in pumping cycles 
without any charge accumulated inside the dot. In this case, we have 
${\hat \phi} ^L = - {\hat \phi}^R$. So we may define 
the circulating current $I$
as $I \equiv Q_c/\tau$, which takes the form
\begin{equation}
 I = \frac {{\hat \phi}^L -{\hat \phi}^R} {4 \pi \tau}.
	\label{current}
\end{equation}
>From this we immediately conclude that 
{\it for an embedded quantum dot with the mirror symmetry, i.e., the 
left-right symmetry, there is no pumping circulating current.} This is 
consistent with Moskalets and B\"uttiker~\cite{mb03} who stated that 
the spatial asymmetry of the scatterer is a necessary condition 
for the existence of an adiabatic pump effect.

We emphasize that Eq.~(\ref{current}) describes the contribution of
a certain discrete energy level to the circulating current. The total
current results from summing up all occupied energy levels, which
depends on the number of electrons in the ring.

\textit{A numerical example.}
Consider a quantum dot modeled by a potential $V(x)$, which is defined 
as $0$ for $|x|\geq a$, $V_1$ for $-a < x < -b$, $V_2$ for $|x| \leq b$, and 
$V_3$ for $b < x < a$. Here $x$ denotes the coordinate along the ring.
The same potential has been used to model the quantum dot embedded in
a double path interferometer proposed to directly observe scattering
geometric phase~\cite{zhou2}.
Then the instantaneous spin-independent $2 \times 2$ scattering matrix
$S(t)$ for the dot may be determined from the solution of 
the Schr\"odinger equation
$(-(\hbar ^2 /2m) \partial ^2/ \partial x^2 + V(x)- E ) \psi =0$.
For a dot of size 800 nm 
the energy level spacing is of the order of 4.5 meV. 
The Coulomb energy, assuming a dielectric constant of 10, 
is of the order of 0.08 meV. 
Thus, the dimension of the dot is such that the Coulomb energy is much
less than the separation between the resonances and can be ignored. 
Also the spin-dependent scattering inside the dot is ignored. 

Suppose we periodically and adiabatically vary three gate voltages
$V_1$, $V_2$, and $V_3$.  That is, we choose independent
external parameters $X^1$, $X^2$ and $X^3$ as the gate voltages
$V_1$, $V_2$, and $V_3$, respectively,  which
allows us to control the scatterer in different ways.
For instance, we can choose to adiabatically change $V_1$ and $V_2$ 
with $V_3$ kept constant, i.e., 
$V_1 = V^0_1 + \Delta V_1 \sin \omega t,
~V_2 = V^0_2 + \Delta V_2 \sin \omega t, V_3 = V^0_3, 
~(\Delta V_{1,2} \ll V^0_{1,2})$, with 
$\omega$ being the slow frequency characterizing the adiabaticity. 

\begin{figure}
\begin{center}
\includegraphics[width=\columnwidth]{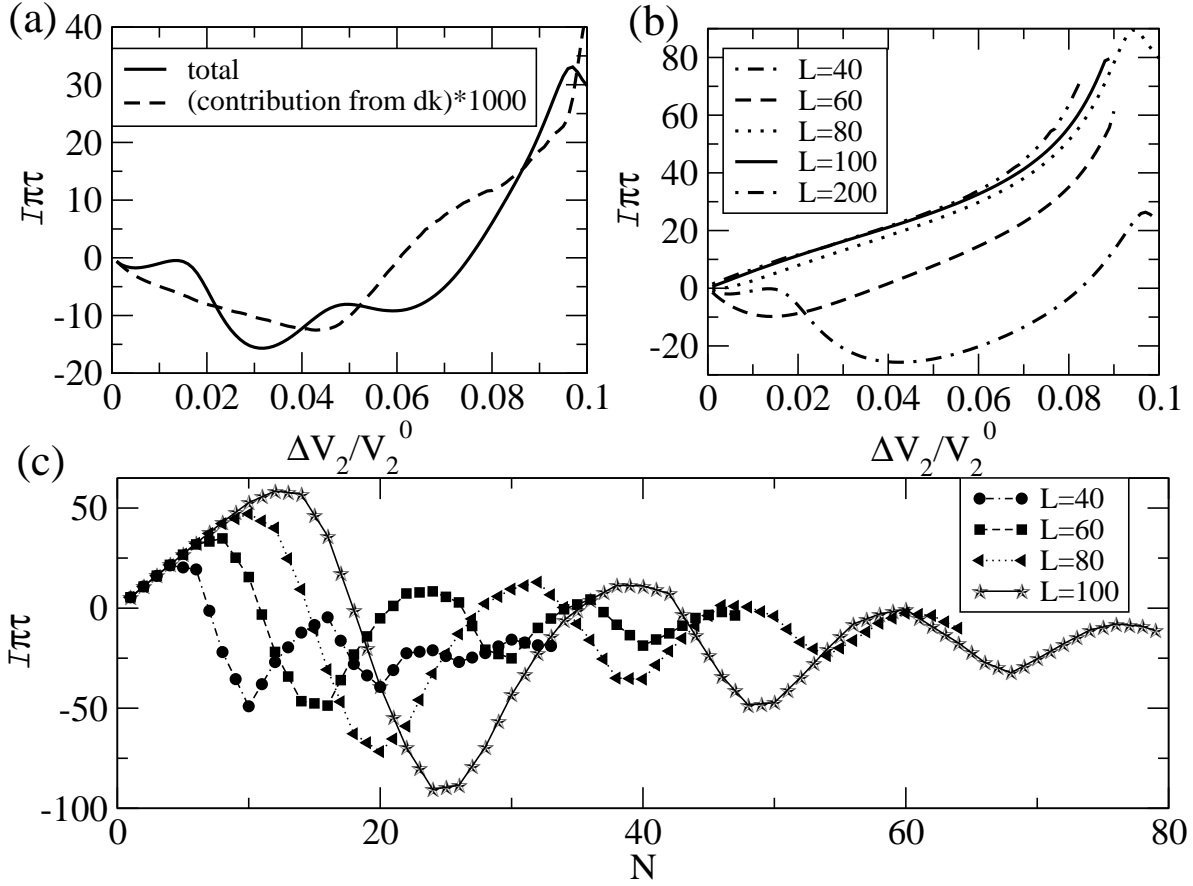}
\end{center}
\caption{\label{result_all:fig}
(a) shows the 
total current as a function of the contour $\Delta V_2/V_2^0$, and 
contribution to the total current from d$k$. 
As we can see the contribution from the time dependence of $k$ is quite small. 
The ring size is 40 (if the size of the QD is 18 in the same arbitrary units), 
and the number of electrons in the ring is 58. \\
(b) shows the 
total current for the first 10 electrons for different sizes of the 
ring as a function of the size of the contour $\Delta V_2/V_2^0$. 
The current increases as a function of the ring-size for a fixed 
number of electrons. \\
(c) shows the 
total current as a function of the number of electrons in the ring, 
for different sizes of the ring. Oscillations in the total current 
is clearly seen. We used $\Delta V_2=0.1V_2^0$, and $\Delta V_1=0.1V_1^0$. 
The size of the QD is 18.}
\end{figure}

The allowed discrete $k$'s may be solved numerically from the dispersion
equation, which results in time-dependent discrete energy levels when
the dot undergoes an adiabatic cycle. 
Suppose there are totally $N_e=2M$ electrons in the ring. Then at low
temperature, they occupy the lowest $M$ level, with two electrons in
each level.  The total current ${\cal I}$ takes
${\cal I}\equiv 2 \sum ^M_{l=1} I_{(l)}= 1/(2\pi \tau) \sum ^M_{l=1}
 ({\hat \phi}^L_{(l)} -{\hat \phi}^R_{(l)})$, with
the subscript $l$ labels the lowest $M$ energy levels.
In Fig.~\ref{result_all:fig}(a), we plot the total current, \IPT defined 
positive clockwise, for 58 electrons
as a function of the parameter $\Delta V_2$. The current is 
most sensitive to this parameter. We also plot the contribution to the 
current from the time dependence of $k$. This contribution is 
quite small. In  Fig.~\ref{result_all:fig}(b) we keep the number of 
electrons constant at 10 and change the size of the ring. 
When the size of the ring 
is large compared to the size of the QD (L=100 and 200 compared to a QD size 
of 18), then, when increasing the size the current increases only 
very little. 
The most interesting behaviour is shown in  Fig.~\ref{result_all:fig}(c). 
Here we plot the total current as a function of the number of 
electrons for different ring sizes. 
Compared to the persistent current\cite{pcurrent} in the ring,
there is no large odd/even effect 
where the current changes sign with the parity. 
However, in this figure we can see that the current can change sign 
(thus direction) as a function of the filling, and 
the effect is larger for a smaller number of 
electrons. This can be detected by measuring the magnetic field 
produced by electrons traveling around the ring. 

In conclusion, we have studied adiabatic scattering in a ring geometry and 
established the connection between the circulating
(pumping) current and the geometric phases acquired by the time-reversed
scattering states based on a pure geometric argument.
This may be viewed as a generalization of persistent currents in a 
mesoscopic ring, in the sense that the Aharonov-Bohm effect is a special
case of Berry's phases. We have also performed numerical calculations on a 
quantum dot embedded in a small metallic ring. 

\subsection{Acknowledgments}
This work was supported by the Australian Research Council.
We thank R.H.\ McKenzie for helpful discussions and supporting this work.


\section*{References}

\begin{thebibliography}{99}

\smallskip
\bibitem{berry}
Berry M V 1984 {\it Proc.\ Roy.\ Soc.\ London, Ser.\ A} {\bf 392} 45
\smallskip

\smallskip
\bibitem{sw89}
Shapere A and Wilczek F 1989 {\it Geometric Phases in Physics}
(Singapore: World Scientific)
\smallskip

\smallskip
\bibitem{jones00}
Jones J A, Vedral V, Ekert A and Castagnoli G 2000
{\it Nature} {\bf 403} 869; 
Falci G, Fazio R, Palma G, Siewert J and Vedral V 2000 
{\it ibid} {\bf 407} 355;
Filipp S and Sj\"oqvist E 2003 \PR {\it Lett.} {\bf 90} 050403; 
Sj\"oqvist E, {\it et al.} 2000 {\it ibid} {\bf 85} 2845; 
Manini N and Pistolesi F 2000 {\it ibid} {\bf 85} 3067
\smallskip

\smallskip
\bibitem{zhou1}
Zhou H-Q, Cho S Y, and McKenzie R H 2003  \PR {\it Lett.} {\bf 91} 186803
\smallskip

\smallskip
\bibitem{b98}
Brouwer P W 1998 \PR {\it B} {\bf 58} 10135
\smallskip
 
\smallskip
\bibitem{mb02}
Moskalets M and B\"uttiker M 2002 \PR {\it B} {\bf 66} 035306;
Zhou F, Spivak B and Altshuler B 1999  \PR {\it Lett.} {\bf 82} 608;
Andreev A and Kamenev A 2000 {\it ibid.} {\bf 85} 1294;
Avron J E, Elgart A, Graf G M and Sadun L 2001 {\it ibid.} {\bf 87} 236601;
Makhlin Y and Mirlin A D 2001 {\it ibid.} {\bf 87} 276803
\smallskip

\smallskip
\bibitem{zhou2}
Zhou H-Q, Lundin U, Cho S Y and McKenzie R H 2004 \PR {\it B} 
{\bf 69} 113308
\smallskip

\smallskip
\bibitem{switkes}
Switkes M, Marcus C M, Campman K and Gossard A C 1999 
{\it Science} {\bf 283} 1905
\smallskip

\smallskip
\bibitem{cohen03}
Cohen D 2003 \PR {\it B} {\bf 68} 201303
\smallskip

\smallskip
\bibitem{mb03}
Moskalets M and B\"uttiker M 2003 \PR {\it B} {\bf 68} 161311
\smallskip

\smallskip
\bibitem{landauer}
Landauer R 1957 {\it IBM J.\ Res.\ Dev.} {\bf 1} 223
\smallskip

\smallskip
\bibitem{btp}
B\"uttiker M, Thomas H and Pr\`etre A 1994 {\it Z.\ Phys.\ B} {\bf 94} 133
\smallskip

\smallskip
\bibitem{mailly93}
Mailly D, Chapelier C and Benoit A 1993 \PR {\it Lett.} {\bf 70} 2020
\smallskip

\smallskip
\bibitem{dtime}
Since the adiabatic scatterings in the dot preserves 
the necessary condition
between the characteristic time scales, i.e.,
$\tau_d \ll \tau_s \ll \tau$,
the dwell time for scattering events can be defined by
$\tau_d = N/J$, where $J$ is the incident flux, $J=\hbar k/m$,
and the number of particles in the dot is given by
$N = \int_{\rm dot} |\psi_{\rm dot}(x)|^2 dx$.
Here, $\psi_{\rm dot}(x)$ is the wave function describing
the dot region.
In our model of ring geometry,
the dwell time is dependent of the discrete $k$'s
due to the boundary condition.

\smallskip
\bibitem{thirring}
Narnhofer H and Thirring W 1982 \PR {\it A} {\bf 26} 3646;
Avron J E, Elgart A, Graf G M and Sadun L 2002 
{\it J.\ Math.\ Phys.} {\bf 43} 3415
\smallskip

\smallskip
\bibitem{pt}
Simon B 1983 \PR {\it Lett.} {\bf 51} 2167;
Aharonov Y and Anandan J 1987 {\it ibid.} {\bf 58} 1593;
Samuel J and Bhandari R 1988 {\it ibid.} {\bf 60} 2339

\smallskip
\bibitem{pcurrent}
Cheung H -F, Gefen Y, Riedel E K and Shih W -H
\PR {\it B} {\bf 37} 6050 (1988)

\end{thebibliography}
\end{document}